# Physical embedding machine learning force fields for organic systems

Junbao Hu[1,2], Dingyu Hou[1,2], Jian Jiang[1,2*]

[1] Beijing National Laboratory for Molecular Sciences, State Key Laboratory of Polymer Physics and Chemistry, Institute of Chemistry, Chinese Academy of Sciences, Beijing 100190

[2] University of Chinese Academy of Sciences, Beijing 100049

*Corresponding Author(s): jiangj@iccas.ac.cn

*Machine Learning, Force Field, Physics embedding, Simulation stability, Macroscopic property prediction*

**ABSTRACT:** Machine learning force fields possess unprecedented potential in achieving both accuracy and efficiency in molecular simulations. Nevertheless, their application in organic systems is often hindered by structural collapse during simulation and significant deviations in the prediction of macroscopic properties. Here, two physics-embedded strategies are introduced to overcome these limitations. First, a physics-inspired self-adaptive bond-length sampling method achieves long-timescale stable simulations by requiring only several tens of single-molecule data sets, and has been validated across molecular systems, including engineering fluids, polypeptides, and pharmaceuticals. Second, a top-down intermolecular correction strategy based on a physical equation is introduced. This strategy requires only a small amount of simulation data and completes the optimization of tunable parameters within a few hours on a single RTX 4090 GPU, significantly reducing errors in density and viscosity, as validated in systems including ethylene carbonate, ethyl acetate, and dimethyl carbonate. Together, these approaches directly integrate physical insights into the machine learning models, thereby enhancing robustness and generalizability, and providing a scalable pathway for physics-embedded machine learning force fields.

## INTRODUCTION

Molecular dynamics (MD) simulation is a fundamental tool for investigating microscopic mechanisms and macroscopic properties in chemistry, molecular biology, physics, and materials science[1-4]. Although ab initio molecular dynamics (AIMD) provides high accuracy, its computational cost is prohibitive for large-scale systems. In contrast, traditional simulation methods based on empirical force fields are computationally efficient but often suffer from insufficient accuracy when applied to complex systems. In recent years, machine learning force fields (MLFFs) have emerged as a promising alternative, offering a favorable balance between accuracy and efficiency. As a result, they are increasingly applied in molecular simulations and have become a research hotspot[5-9].

Despite notable progress, applying MLFFs to organic molecular systems remains challenging, particularly for long-timescale simulations. Organic systems simultaneously involve strong intramolecular interactions (e.g., covalent bonds) and weak intermolecular interactions (e.g., van der Waals forces, hydrogen bonding, π–π stacking), which differ by orders of magnitude in energy scale. Intramolecular interactions govern conformational changes and vibrational dynamics, whereas intermolecular interactions, though weaker, critically determine macroscopic properties such as density and viscosity. Inadequate modeling of either type of interaction can result in structural collapse (e.g., bond breaking, atomic overlap[10-12]) or failure in predicting macroscopic properties (e.g., density) [13].

Structural collapse during molecular simulations is typically attributed to prediction errors in high-energy bond-length regions, especially when the training set lacks coverage of relevant configurations. Research has shown that even a single chemical bond prediction error can destabilize the entire simulation[14] [10, 15, 16]. To address this issue, several improvement strategies have been proposed. Wang[16] enhanced simulation stability by introducing novel loss functions, such as maximum force error, mass-normalized error, and conformational cluster force error. However, these methods require large training datasets. Cui[14] proposed a two-stage self-supervised learning approach combined with test-time adaptive fine-tuning. This method improved stability but required encoder fine-tuning at each simulation step, thereby resulting in high computational complexity and limiting its applicability to long-term simulations of large-scale systems. Our previous work[15] partially addressed the issue of structural collapse by enhancing dataset coverage in high-energy bond length regions through a dynamic bond stretching method.. However, this approach was validated only in simple systems (e.g., perfluorotributylamine) , and it would be considerably difficult to extend to more complex organic molecules. Complex organic molecules often contain diverse bond types (single, double, triple bonds) and ring structures (e.g., benzene rings), making uniform stretching factors unsuitable. For instance, the permissible elongation range of a carbon–carbon triple bond is much smaller than that of a single bond. Excessive stretching can cause self-consistent field (SCF) non-convergence or produce anomalous atomic forces, thereby increasing model training difficulty.

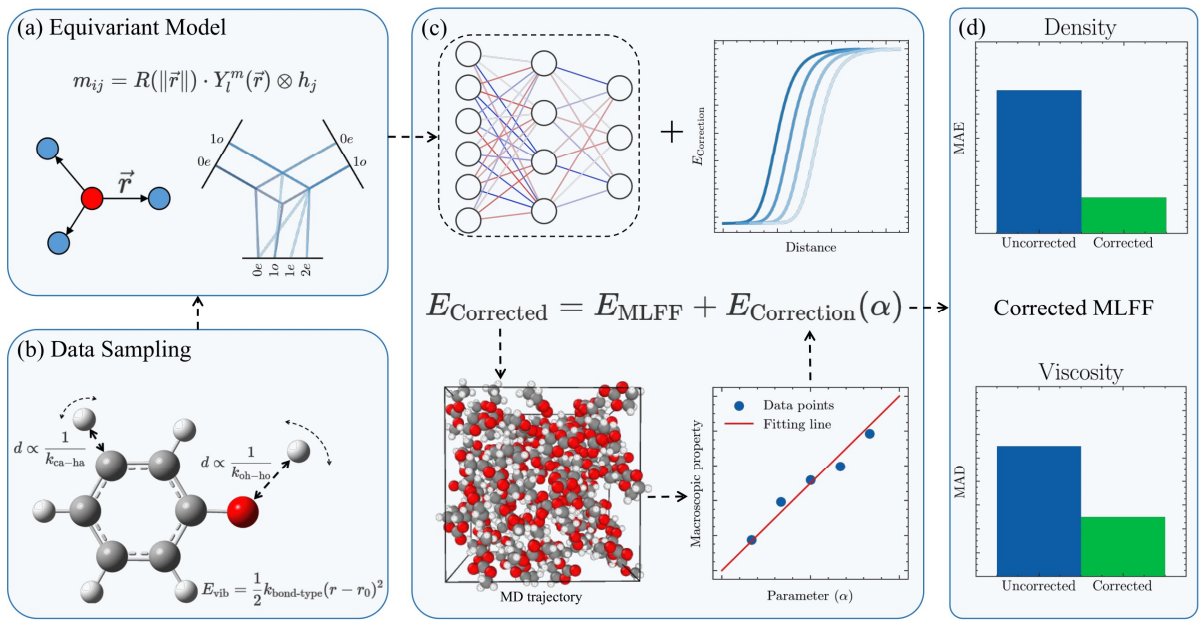

**Figure 1.** Schematic workflow of physics-embedded machine learning force fields. (a) An equivariant framework is adopted as a physical symmetry constraint and embedded into the MLFF model. Here, $m$, $R$, $Y$ and $h$ denote interatomic messages, radial basis functions, spherical harmonics, and atomic features, respectively, while $i$ and $j$ represent different atoms. (b) A self-adaptive bond length sampling method is constructed based on physical knowledge. Here, $d$, $k$, and $E$ denote bond length, force constant, and bond potential energy, respectively. (c) A top-down intermolecular interaction correction method based on the embedding of physical equations. Here, $E$ represents energy, with subscripts used to distinguish different energy categories, and $\alpha$ represents an optimizable parameter. (d) Performance of the corrected MLFF in predicting macroscopic properties such as density and viscosity.

Manually setting stretching factors for each chemical bond is extremely cumbersome. These limitations highlight the urgent need for a more generally applicable and physically informed improvement method.

A second major challenge arises from the disparity between intramolecular and intermolecular interactions. Because intramolecular interactions are substantially stronger, the optimization process during loss minimization often prioritizes fitting them, leaving intermolecular interactions neglected. As a result, MLFFs may achieve low average force errors (e.g., < 60 meV/Å[17, 18]) while still producing large deviations in macroscopic property predictions. This is because the variation in system energy is dominated by intramolecular interactions, which prevents the intermolecular interactions from being adequately fitted. However, most macroscopic properties such as density, viscosity, and dielectric constant are highly sensitive to intermolecular interactions[13]. Accordingly, recent efforts have focused on developing advanced approaches to address this issue. For example, the ByteDance team[19] introduced strategies such as multi-round distillation, ensemble learning, and density alignment to improve property prediction. However, their approach depends on repeated fine-tuning and extensive simulation data, resulting in a complex workflow with limited reproducibility.

In summary, purely data-driven MLFF models face inherent limitations in achieving accurate descriptions of both intramolecular and intermolecular interactions. Physics embedding has therefore emerged as a key pathway for enhancing the robustness and transferability of MLFFs. To address these challenges, this work introduces three physics-embedded strategies for MLFFs, encompassing data sampling, model symmetry constraints, and post-training correction. Among them, model symmetry constraints are represented by high-order equivariant models, which have already demonstrated excellent performance across a broad range of tasks[10, 11, 20, 21]. Consequently, the present work primarily focuses on the other two complementary strategies: data sampling and post-training correction. The first is a physics-inspired self-adaptive bond-length sampling method. By incorporating empirical physical knowledge into the data sampling stage, this approach adaptively defines bond-length modification ranges and sampling probabilities, thereby enhancing the stability of MLFF simulations. The second is a top-down intermolecular correction by embedding physical equations, which leverages a small amount of simulation data to reduce errors in the prediction of macroscopic properties, such as density and viscosity. These two approaches respectively address the issues of intramolecular structural collapse and inadequate fitting of intermolecular interactions, and collectively offer complementary enhancements that significantly improve both the accuracy and the physical fidelity of MLFFs. The overall workflow is schematically illustrated in Fig. 1. Both methods can be seamlessly integrated into existing MLFF frameworks directly, while demonstrating strong adaptability and robustness across diverse molecular systems.

## RESULTS AND DISCUSSION

### Self-adaptive Bond Length Sampling Method Inspired by Physical Knowledge

This study proposes a self-adaptive bond length sampling strategy that incorporates physical knowledge to refine the molecular structure of the original dataset. Based on our previous dynamic bond stretching method[15], in this work, we propose a self-adaptive bond length sampling strategy by integrating topological information from empirical force fields to distinguish chemical bond types based on physical rules. It adaptively sets bond length modification ranges and sampling weights using physical parameters, thereby embedding domain knowledge from empirical force fields into the data sampling process and effectively enhancing the simulation stability of MLFF. The procedure consists of the following four main steps:

**Bond type classification using physical rules.** Topological files from empirical force fields (e.g., GAFF[22] or OPLS[23]) are used to systematically obtain atom types and automatically categorize chemical bonds. For example, in paracetamol (Fig. 2a), although only four elements (C, N, O, H) are present, GAFF distinguishes 10 atom types based on chemical environment. For example, hydrogen atoms bonded to aliphatic carbon, aromatic carbon, nitrogen, and oxygen are labeled as "hc," "ha," "hn," and "ho," respectively. Furthermore, this classification can effectively differentiate between carbon-oxygen single bonds and double bonds, as well as between carbon-carbon single bonds and aromatic carbon-carbon bonds, thereby realizing classification based on physical knowledge.

**Bond length modification using physical parameters.** This strategy determines self-adaptive bond length sampling ranges using molecular bond length distributions and force field topological files. The modification range for each bond type is divided into a compression segment $[r_{min} \times s_{compression}, r_{min}]$ and a stretching segment $[r_{max}, s_{stretch} \times (k_{ref}/k) + r_{mean}]$. Here, $r_{min}$, $r_{mean}$ and $r_{max}$ represent the minimum, mean, and maximum values of the bond length of the data set, respectively; $k$ is the bond force constant from the topological file; $k_{ref}$ is the force constant of a reference bond (typically "c3-c3"); $s_{compression}$ is set to 0.95 to avoid unphysically short bond lengths; and $s_{stretch}$ is usually set to 0.25 to avoid SCF convergence difficulties caused by excessive stretching. Sampling probabilities are weighted by the normalized reciprocal of bond force constant, thereby ensuring that unstable bonds exhibit a broader stretching range and higher sampling probabilities. The sampling ratio between compression and stretching segments is set to 1:4 to enhance sampling of stretched configurations.

**Tailored modification of terminal groups.** Since most structural collapses originate from terminal bonds, for those bonds involving terminal atoms (e.g., hydrogen or halogens), the bond force constant $k$ is multiplied by 0.8 to increase stretching amplitude and sampling frequency. As a consequence, terminal groups are randomly rotated by 15~30° around the bond axis with a probability of 0.8 to enrich structural diversity.

**Post-modification geometry check.** To avoid abnormal molecular structures, for stretching modifications, if the modified structure contains any non-bonded atom pairs that are unphysically close (i.e., when the interatomic distance is shorter than the equilibrium bond length by less than 0.25 Å), the modification is discarded, and another pair of atoms is selected for modification until a valid modification is obtained.

Compared with our previous work, this strategy offers four advantages: (i) systematic bond classification using empirical topological information; (ii) self-adaptive adjustment of bond stretching (or compression) ranges and sampling probabilities according to bond force constants; (iii) specific treatment of terminal groups; and (iv) explicit geometric constraints ensuring structural plausibility.

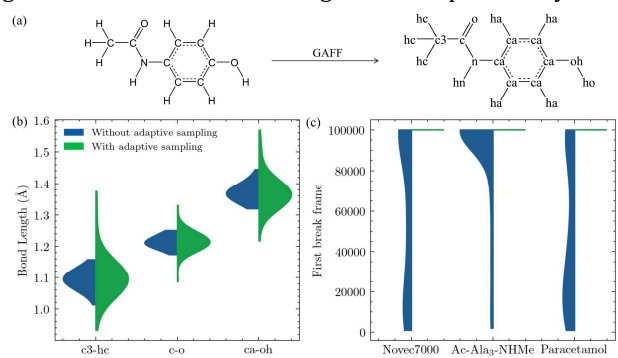

**Figure 2.** (a) Comparison between molecular structures annotated by element types and those annotated by topological information from empirical force fields, using paracetamol as an example; (b) Comparison of the original bond length distribution and the distribution after self-adaptive bond length sampling; (c) Stability test results of molecular dynamics simulations for different molecular systems. Blue and green indicate results from the original method and those with self-adaptive bond length sampling, respectively.

In validation experiments, three representative organic molecules are selected: Novec7000 ($C_4H_3F_7O$, an engineering fluid), Ac-Ala$_3$-NHMe ($C_6H_{12}N_2O_2$, a peptide), and paracetamol ($C_8H_9NO_2$, a drug). When conducting stability tests based on the MACE[24] model, each molecule is trained using five different random seeds. All models are used for subsequent tests to reduce the interference caused by randomness. Molecular simulation stability tests are carried out using 20 randomly selected samples from the trajectory structures generated by the GAFF empirical force field as the initial structures, with a temperature of 800 K and an integration step of 1 fs for 100 ps of MD simulation. Therefore, a total of 100 independent trajectories are obtained for each molecule. High-temperature conditions are chosen to enhance sampling efficiency in MD, enabling effective evaluation of model stability within limited simulation time. Structural collapse is defined as either atomic overlap or bond lengths exceeding 0.5 Å from equilibrium. If no collapse occurred throughout the entire simulation, the model is considered stable, and the final frame number is recorded. The simulation stability results for each molecule are presented in Fig. 2c.

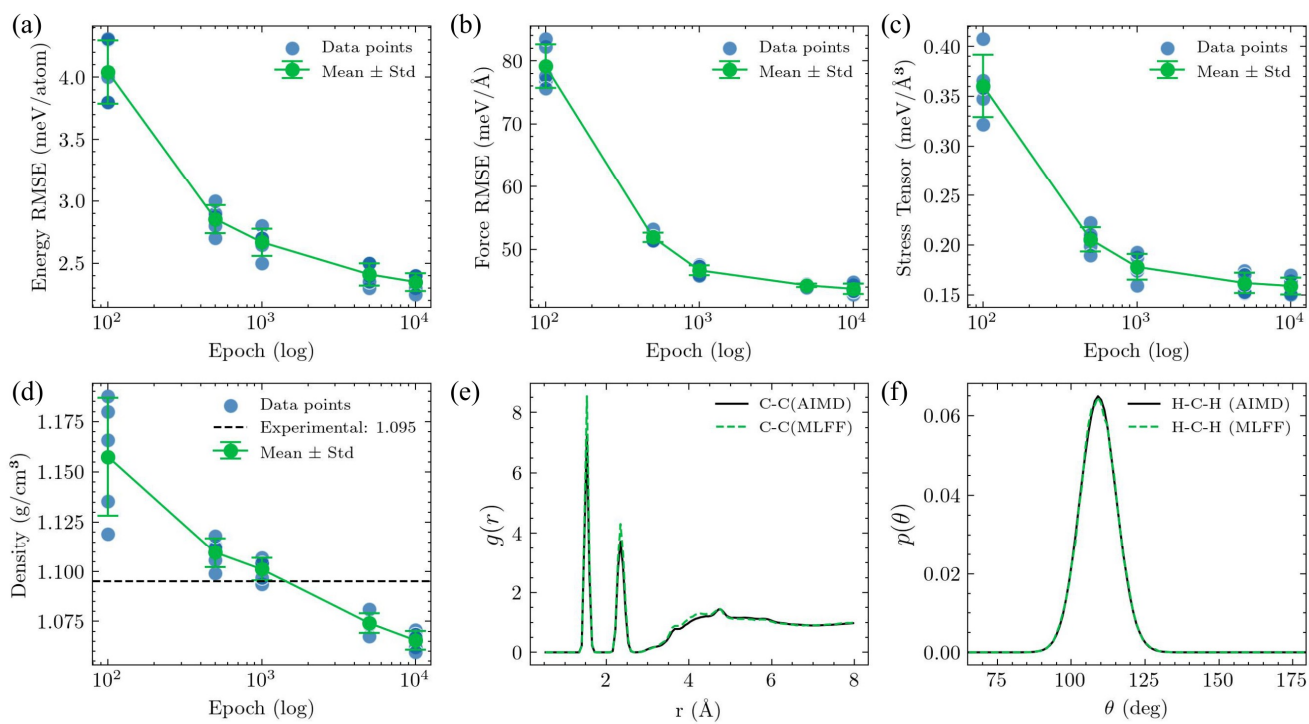

**Figure 3.** (a–d) Evolution of predicted energy, force, stress tensor, and density values from MLFF with training epochs. Blue points represent individual samples, green points denote mean values, error bars indicate standard deviations, and black dashed lines correspond to experimental density values; (e, f) Comparison of radial distribution function (RDF) and angular distribution function (ADF) obtained from MLFF after 100 training epochs with AIMD results. Black solid lines denote AIMD results, and green dashed lines denote MLFF results.

Despite the strong generalization ability and data efficiency of the MACE model, when using only 50 training samples, the structural collapse probabilities of Novec7000, Ac-Ala$_3$-NHMe, and paracetamol still reached 59%, 22%, and 77% respectively. To test the proposed strategy, 45 structures are modified using the self-adaptive sampling method, with 5 samples reserved for validation. For example, in paracetamol, the C–O bond (labeled "ca-oh" in GAFF force field topological files) and C=O bond ("c-o") are stretched by 0.20 Å and 0.12 Å beyond equilibrium, respectively, while the C–H bond ("c3-hc") is stretched by 0.29 Å beyond equilibrium (Fig. 2b). These results illustrate the method's ability to adjust sampling adaptively across different bond types. As shown in Fig. 2c, after enhancement with this strategy, the MACE model successfully passed all 100 stability tests across the three systems, demonstrating substantial improvements in MLFF stability and robust adaptability across diverse molecular structures.

### Top-Down Intermolecular Correction Based on Embedded Physical Equation

We first examined the discrepancies between microscopic accuracy and macroscopic property prediction of MLFFs across different training epochs. A benchmark dataset for a mixed solvent of ethylene carbonate (EC) and ethyl methyl carbonate (EMC), computed at the PBE-D3(BJ)/TZVP level, is used for evaluation. This dataset, constructed via a continual learning strategy, contains over 900 samples combining rigid volume scans with various component ratios, thereby providing comprehensive coverage of the configurational space of intermolecular interactions.

The MACE model (denoted MACE-EC/EMC) is trained in multiple independent rounds with different random seeds. As shown in Fig. 3, prediction errors for energy, forces, and stress tensor gradually decreased with training. At epoch 100, the root mean square error (RMSE) of forces reached approximately 80 meV/Å, and the radial distribution function (RDF) and angular distribution function (ADF) closely matched AIMD results. Despite the high accuracy achieved for microscopic properties such as forces and structures, the predictions of macroscopic density at epoch 100 exhibited significant fluctuations, with an average value deviating significantly from the experimental benchmark. This discrepancy indicates inadequate fitting of intermolecular interactions.

As training progressed, predicted densities gradually decreased but did not converge even after 10,000 epochs. At this stage, the predicted density is notably lower than experimental values, with large variations across different random seeds, highlighting the difficulty of accurately capturing intermolecular interactions and the poor reproducibility of the model. Interestingly, the value of the density predicted at epoch 1000 is close to experiment, but this agreement likely resulted from a fortuitous cancellation of systematic biases and random errors rather than physical accuracy.

To further examine the influence of the reference model, we evaluated the foundation model MACE-OFF23(S) [25], trained on ωB97M-D3(BJ)/def2-TZVPPD data. Although this model also employed the MACE architecture and required approximately six days of training, its predicted density and viscosity still significantly exceeded experimental values (see Table 1 and Table 2).

These findings suggest that the pronounced deviations of MLFF-based molecular simulations in predicting the macroscopic properties of organic systems are twofold in origin: (i) a bias in the fitting of intermolecular interactions arising during MLFF training, especially when undertrained; and (ii) systematic errors inherent to the training datasets itself, linked to the choice of functional and basis set of quantum method.

To mitigate these issues, we propose a top-down correction strategy that selectively enhances intermolecular interactions while minimally perturbing intramolecular contributions. We adopt the widely used DFT-CSO dispersion correction equation[26]:

$$E_{\text{DFT-CSO}} = -\sum_{AB}\left[s_6 + \frac{a_1}{1+\exp(R_{AB}-2.5R_0^{AB})}\right]\frac{C_6^{AB}}{R_{AB}^6+(2.5^2)^6}$$

where $C_6^{AB}$ is the dispersion coefficient, $R_{AB}$ is the interatomic distance, $R_0^{AB}$ is the equilibrium distance; $a_1$ and $s_6$ are damping function parameters.

The advantages of this equation are: its concise and universal form, allowing flexible adjustment of dispersion strength primarily by tuning $a_1$;[26] clear physical meaning, describing purely attractive intermolecular interactions; and minimal impact on intramolecular interactions.

Distinct from traditional micro-level dispersion corrections based on high-level theory, this work adopts a top-down strategy aiming at experimental macroscopic values. Furthermore, this correction strategy is capable of modulating intermolecular attractive or repulsive effects, as MLFFs may either underestimate or overestimate intermolecular attraction. Therefore, the constructed correction energy term takes the form:

$$E_{\text{corrected}} = E_{\text{MLFF}} + \alpha \cdot E_{\text{DFT-CSO}}$$

where $E_{\text{MLFF}}$ is the original MLFF energy, and $\alpha$ is a scaling factor tuning the strength and direction of correction (positive values strengthen attraction; negative values enhance repulsion). Experiments show that even extremely small values of $\alpha$ (such as ±0.05) can significantly improve the prediction results.

We selected density as the correction target because of its strong dependence on intermolecular interactions, low computational cost, rapid convergence, and reliable experimental accessibility. Density optimization is carried out at 298 K for two systems: (i) the EC/EMC (3:7 mass ratio) mixture using MACE-EC/EMC, and (ii) pure EMC using MACE-OFF23(S). As shown in Fig. 4a and Fig. 4b, density displays a clear linear relationship with the parameter $a_1$, allowing the optimal parameter to be determined with only a few samples. The best-performing values are $a_1 = 0.6$ ($\alpha = 0.05$) for MACE-EC/EMC and $a_1 = 1.07$ ($\alpha = -0.05$) for MACE-OFF23(S). It is worth noting that this correction task can be accomplished within five hours on a RTX 4090 GPU. The strong physical constraints significantly reduce its reliance on simulation data, yielding exceptional data efficiency and practical applicability.

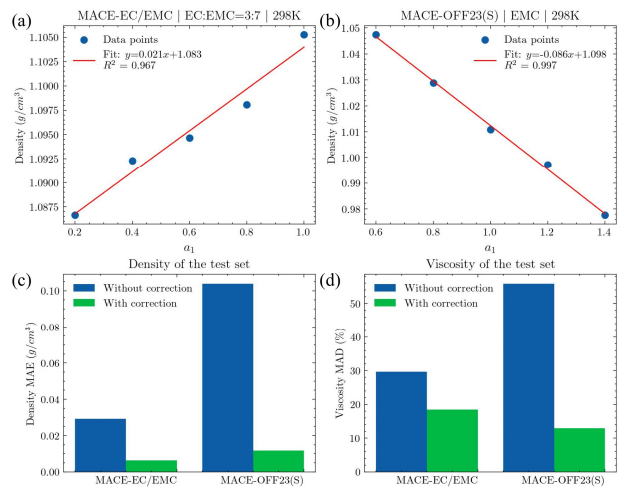

**Figure 4**. (a, b) Intermolecular corrections of MACE-EC/EMC and MACE-OFF23(S) models in an EC/EMC mixture with a 3:7 mass ratio and in pure EMC solvent, respectively. The horizontal axis $a_1$ denotes the tuning parameter, and the vertical axis denotes density. Red lines represent fitted curves, and blue points denote sampled data; (c, d) Density and viscosity prediction results of MACE-EC/EMC and MACE-OFF23(S) models on the test set. Blue and green represent results with correction and without correction, respectively.

The generalization capability of the proposed method is further evaluated across different temperatures and systems, including EC, EMC, ethyl acetate (EA), and dimethyl carbonate (DMC). As shown in Figure 3(c, d), the mean absolute error (MAE) of density predictions on the test set for the corrected MACE-EC/EMC and MACE-OFF23(S) models decreased by 78% and 88%, respectively, reaching 0.006 g/cm³ and 0.012 g/cm³. Similarly, the mean absolute deviation (MAD) for viscosity predictions decreased by 38% and 77%, with final deviations from the experimental values being 18.4% and 12.9%, respectively. These improvements markedly outperform the uncorrected models, with prediction errors remaining within experimentally acceptable ranges (density error < 0.01 g/cm³, viscosity fluctuation 20%)[19]. Moreover, Table 2 shows that corrected performance is comparable to or even exceeds that of the Bamboo model[19]. To assess the influence of this correction strategy on intramolecular interactions, Fig. 5 presents the performance of MACE-EC/EMC and MACE-OFF23(S) on rigid-volume scanning test sets for their respective correction systems (a 3:7 mass-ratio EC/EMC mixed solution and pure EMC solvent). The difference in force RMSE before and after correction is negligible (<0.8 meV/Å), which is much lower than the model fitting limit (about 44 meV/Å) shown in Figure 2(b). This result also indicates that intermolecular interactions are difficult to optimize through conventional training and instead require post-training top-down correction. The shifts in energy minima observed in Figs. 4(a, d) suggest that the correction enhances intermolecular attraction in MACE-EC/EMC and intermolecular repulsion in MACE-OFF23(S), thereby improving the interpretability of MLFFs as otherwise black-box models. Furthermore, Figs. 4(c, f) demonstrate that the correction has almost no effect on radial distribution functions (RDFs).

Table 1 | Comparison results of the density of MACE-EC/EMC with and without correction. The underline indicates the top-down density correction system.

| Model | System | Temperature(K) | Density(g/cm$^3$) | | |
|---|---|---|---|---|---|
| | | | Without correction | With correction | Experiment[19, 27] |
| MACE-EC/EMC | EC | 313 | 1.278 | 1.308 | 1.321 |
| | EMC | 298 | 0.986 | 1.017 | 1.006 |
| | EC:EMC=5:5 | 298 | 1.125 | 1.153 | 1.157 |
| | EC:EMC=3:7 | 293 | 1.072 | 1.101 | 1.101 |
| | EC:EMC=3:7 | 298 | 1.067 | 1.095 | 1.095 |
| | EC:EMC=3:7 | 303 | 1.063 | 1.093 | 1.089 |
| | EC:EMC=3:7 | 313 | 1.053 | 1.084 | 1.078 |
| MACE-OFF23(S) | EC | 313 | 1.400 | 1.325 | 1.321 |
| | EMC | 298 | 1.117 | 1.004 | 1.006 |
| | EA | 298 | 1.025 | 0.922 | 0.900 |
| | DMC | 298 | 1.168 | 1.069 | 1.060 |

Table 2 | Comparison results of the viscosity of MACE-OFF23(S) with and without correction.

| Model | System | Temperature(K) | Viscosity (cP) | | | |
|---|---|---|---|---|---|---|
| | | | Without correction | With correction | Bamboo | Experiment[19] |
| MACE-EC/EMC | EC | 313 | 1.01 ±0.041 | 1.20 ±0.060 | 1.38 | 1.60-1.90 |
| | EMC | 298 | 0.54 ±0.006 | 0.65 ±0.012 | 0.72 | 0.65 |
| | EC:EMC=3:7 | 293 | 0.72 ±0.022 | 0.84 ±0.018 | 0.84 | 1.11 |
| | EC:EMC=3:7 | 303 | 0.71 ±0.017 | 0.80 ±0.025 | 0.84 | 0.97 |
| | EC:EMC=3:7 | 313 | 0.65 ±0.017 | 0.69 ±0.023 | 0.73 | 0.87 |
| MACE-OFF23(S) | EC | 313 | 2.07 ±0.046 | 1.48 ±0.085 | 1.38 | 1.60-1.90 |
| | EMC | 298 | 1.06 ±0.047 | 0.54 ±0.003 | 0.72 | 0.65 |
| | EA | 298 | 0.82 ±0.034 | 0.44 ±0.023 | 0.54 | 0.43 |
| | DMC | 298 | 0.98 ±0.019 | 0.55 ±0.010 | 0.63 | 0.585,0.706 |

Overall, the top-down correction method based on the embedding of physical equations not only effectively enhances the performance of MLFFs for predicting macroscopic properties, but also exerts minimal interference on intramolecular structural behavior.

### Datasets and Model Training

For the Novec7000, Ac-Ala$_3$-NHMe, and paracetamol systems, 1 ns MD simulations are performed with the GAFF[22] force field at 300 K using a 1 fs integration step. Trajectories are saved every 100 steps, and 50 configura-tions are randomly selected as initial samples. Energies and forces are computed using the GFN2-xTB[28, 29] method. The EC/EMC mixed solvent dataset, rigid volume scan data, and corresponding AIMD data are directly cited from the literature[13] without modification.

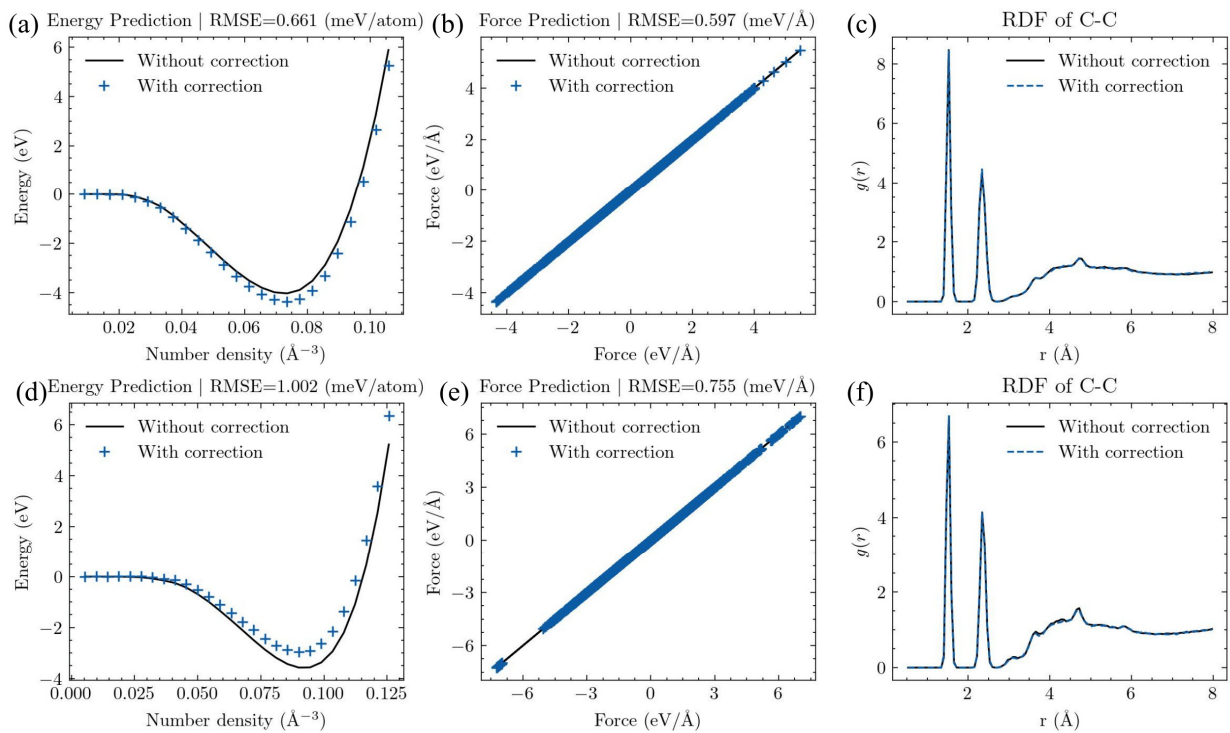

**Figure 5.** Comparison of MLFFs with and without physical equation correction in rigid volume scanning tests. (a, b) Energy and force of MACE-EC/EMC in an EC/EMC mixture with a 3:7 mass ratio; (d, e) Energy and force of MACE-OFF23(S) in pure EMC solvent; (c, f) RDF comparison for the corresponding systems. Black and blue represent results with correction and without correction, respectively.

In this study, the MACE model is adopted as the architectural basis, with GPU-accelerated training implemented using the cuEquivariance library.

Hyperparameter settings for training models of Novec7000, Ac-Ala$_3$-NHMe, and paracetamol are as follows: node feature dimension of 256, maximum expansion order of 3; loss function as the weighted sum of energy and forces with initial weights of 1.0 and 100.0, respectively; batch size of 10; optimizer as Adam with an initial learning rate of 1e-2.

Hyperparameters for the EC/EMC system are as follows: node feature dimension of 64, maximum expansion order of 2; loss function as the weighted sum of energy, forces, and stress tensors with initial weights of 1.0, 100.0, and 100.0, respectively. Stochastic Weight Averaging (SWA) is enabled when training reached 75% of the total epochs, and the weights are adjusted to 100.0, 1.0, and 1000.0; batch size of 32; optimizer as Adam with an initial learning rate of 1e-2.

The foundation model MACE-OFF23(S) is employed without retraining or fine-tuning, with corrections applied solely to intermolecular interactions.

### Molecular dynamics settings

Empirical force field simulations are carried out in OpenMM[30], while MACE-based MLFF simulations are executed in LAMMPS[31] through the ML-IAP interface.

The calculation of density and viscosity followed a two-step workflow. First, a 1 ns NPT simulation (1 fs time step) is conducted to equilibrate the system, and the average density is obtained from the last 200 ps of the trajectory. Subsequently, three configurations are randomly extracted from this trajectory as initial states for 5 ns NVT simulations. Viscosity is then evaluated using the Green–Kubo relation, by time-integrating the autocorrelation function of the off-diagonal components of the pressure tensor.

### CONCLUSIONS

This study proposes two physical embedding strategies to improve the structural stability and macroscopic property prediction accuracy of MLFFs in long-time molecular dynamics simulations.

At the intramolecular interaction level, a self-adaptive bond length sampling method inspired by physical knowledge is proposed. This method distinguishes different chemical bonds using empirical physical rules derived from the topological information of force fields, and adaptively adjusts the sampling range and strategy by incorporating physical parameters such as bond force constants. In this way, this method increases sampling coverage for unstable bonds while preserving conformational validity through geometric constraints, thereby systematically embedding physical knowledge during data generation. The results demonstrate that this approach significantly improves the stability of molecular dynamics simulations. In various organic systems such as engineering solvents, polypeptides, and drug molecules, only 50 single-molecule samples are required to pass rigorous stability tests, indicating that the rational incorporation of physical knowledge can effectively enhance model robustness under small-dataset conditions. It is recommended to use data generated by this method as the initial training set and combine it with active learning[32-34,35] strategies to further expand the dataset, enabling efficient automated sampling of non-bonded interactions such as hydrogen bonding.

At the intermolecular interaction level, a top-down correction strategy based on physical equation embedding is proposed. This approach relies on the DFT-CSO framework, using experimental density as the optimization target, and reduces macroscopic property prediction deviations by tuning a single damping parameter. Benefiting from strong physical constraints, this method is highly data-efficient, with each correction typically completed within a few hours on a single RTX 4090 GPU, demonstrating high efficiency and excellent practical applicability. Across multiple organic solvent test systems, the MAE of density predictions is reduced by 78%~88%, while the MAD of viscosity predictions decreased by 38%~77%. Compared with more complex strategies such as distillation, assembling, and density alignment employed in models like Bamboo, our method achieves comparable or even superior performance with significantly lower computational cost. Furthermore, it can be readily integrated into existing MACE models (including foundation models) via a straightforward parameter optimization process, ensuring strong practicality.

In summary, by embedding physical knowledge into the data sampling stage and integrating physical equations into the model post-processing stage, this work successfully addresses the issues of structural collapse and low accuracy in macroscopic property prediction in traditional MLFF-based molecular simulations, while enhancing the interpretability of the MLFF model.

The concept of physical embedding thus provides a scalable and generalizable pathway for the development of MLFFs, laying the groundwork for high-precision, robust, and transferable molecular simulation tools. Future work will focus on extending the set of adjustable parameters in physical equations (e.g., dispersion coefficient $C_6^{AB}$) to further improve adaptability and generalization in complex systems. Additionally, more efficient calculation methods for macroscopic properties need to be developed, especially for properties requiring long-time sampling, such as viscosity, to reduce the cost of parameter optimization.

## AUTHOR INFORMATION

### Corresponding Author

jiangj@iccas.ac.cn

### Author Contributions

The manuscript was written through the contributions of all authors.

## ACKNOWLEDGMENT

This research was supported by the National Natural Science Foundation of China (Grant Number 22422308).

## ABBREVIATIONS

MD, molecular dynamics; MLFFs, machine learning force fields; AIMD, ab initio molecular dynamics; RDF, radial distribution function; ADF, angular distribution function; EC, ethylene carbonate; EMC, ethyl methyl carbonate; EA, ethyl acetate; DMC, dimethyl carbonate.

Table of Contents

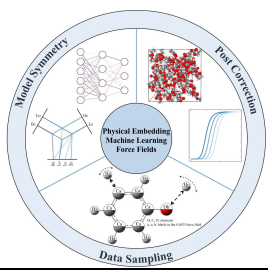